\documentclass{aa}  
\usepackage{graphicx}
\usepackage{txfonts}
\usepackage{siunitx}
\usepackage{soul}
\usepackage[dvipsnames]{xcolor}
\DeclareSIUnit{\angstrom}{\textup{\AA}}
\usepackage[colorlinks=true, linkcolor=blue, citecolor=blue, filecolor=blue, urlcolor=blue]{hyperref}

\begin{document}

    \title{ Imaging and spectropolarimetric observations of a band-split type II solar radio burst with LOFAR }

    \author{S. Normo\inst{1}
            \and
            D. E. Morosan\inst{1}
            \and
            P. Zhang\inst{2,3}
            \and
            P. Zucca\inst{4}
            \and
            R. Vainio\inst{1}}

    \institute{Department of Physics and Astronomy, University of Turku, 20014, Turku, Finland\\
               \email{sanna.l.normo@utu.fi}
               \and    
    Center for Solar-Terrestrial Research, New Jersey Institute of Technology, Newark, NJ, USA
               \and
    Cooperative Programs for the Advancement of Earth System Science, University Corporation for Atmospheric Research, Boulder, CO, USA
               \and
    ASTRON – the Netherlands Institute for Radio Astronomy, Oude Hoogeveensedijk 4, 7991 PD Dwingeloo, The Netherlands
              }    

    \date{Received; accepted}

    \abstract
    {Type II solar radio bursts are generated by electrons accelerated by coronal shock waves. They appear in dynamic spectra as lanes drifting from higher to lower frequencies at the plasma frequency and its harmonic. These lanes can often be split into two or more sub-bands that have similar drift rates. This phenomenon is called band-splitting, and its physical origins are still under debate.}
    {Our aim is to investigate the origin of band-splitting using novel imaging and spectropolarimetric observations of a type II solar radio burst from the Low Frequency Array (LOFAR). }
    {We used LOFAR imaging at multiple frequencies and time steps to track the locations of the radio sources corresponding to the two components of the band-split emission lane. In addition, we estimated the degree of circular polarisation (dcp) for both components using LOFAR's full Stokes dynamic spectra. }
    {From the imaging of the type II burst, we found two close but clearly separated emission regions clustered over several frequencies spanning each split band. One emission region corresponds to the lower frequency band and the other to the higher frequency band of the split lane. Using the full Stokes dynamic spectra, we also found the dcp to be very similar for both bands. }
    {The two distinct emission regions suggest that the split bands originate from two separate regions at the shock. The similar values of dcp for both sub-bands correspond to similar values of magnetic field strength in the two regions and indicate little to no change in the emission region plasma. Thus, our findings are in contradiction with previous theories, which have suggested that split bands originate in the same region but upstream and downstream of the shock. Instead, our results suggest that both bands originate in two separate upstream regions since we find a clear separation in locations and no magnetic compression.}


    \maketitle
    \nolinenumbers
    
    \section{Introduction}   

    Type II solar radio bursts are generally interpreted as radio signatures of coronal shock waves \citep[e.g.][]{Nelson1985}. Fast electrons accelerated by the shock excite Langmuir waves, which are eventually converted into radio emission at the fundamental and/or harmonic of the local plasma frequency \citep{Wild1954,Nelson1985,Cairns2003}. Type II bursts are characterised in a dynamic spectrum as lanes of emission drifting slowly towards lower frequencies. Often there are two lanes of emission with a frequency ratio of approximately two that correspond to emission at the fundamental and at the harmonic of the plasma frequency. In addition to their general spectral appearance, type II bursts can exhibit fine structures such as herringbones. Herringbones are short duration bursts, and they drift towards higher and lower frequencies with a faster drift rate compared to the underlying type II burst \citep{Roberts1959,Cairns1987}. Observational evidence suggests that herringbones originate from individual electron beams accelerated by shock waves driven by coronal mass ejections (CMEs) \citep{Mann2005,Carley2013,Zucca2018,Morosan2019,Zhang2024a}. Another feature encountered in type II bursts is the emission lanes splitting into two or more \citep[see e.g.][]{Zimovets2015,Magdalenić2020} sub-bands, which is known as band-splitting \citep{Vrsnak2001}. However, the physical mechanisms behind band-splitting remains a topic of discussion. 

    From all the proposed mechanisms, there are two theories explaining band-splitting that stand out. The first is a band-splitting model by \cite{Smerd1974,Smerd1975} where the sub-bands correspond to emission on either side of the shock front: one upstream (ahead) and one downstream (behind) of the shock. In this scenario, the lower frequency component (LFC) of the split emission band originates upstream of the shock, and the higher frequency component (HFC) originates downstream of the shock. Since the thickness of the shock front is negligible, the two radio sources are expected to be nearly co-spatial. The upstream-downstream model has been used to infer, for example, the local magnetic field strength and shock properties, such as Mach numbers and density compression ratios, in numerous studies \citep[see e.g.][]{Vrsnak2002,Zimovets2012,Mahrous2018,Zucca2018,Kumari2017,Kumari2019}. The second band-splitting model takes a slightly different approach. \cite{Holman1983} proposed a model where both emission sources originate upstream of the shock but at two different locations along the shock front. The two radio sources are thus expected to be separated in space, and the properties of the split bands cannot be used to determine the above-mentioned properties of the shock. Observational evidence for both the upstream-downstream scenario \citep{Vrsnak2001,Zimovets2012,Zucca2018,Chrysaphi2018,Ramesh2022b} as well as the scenario where both sources originate upstream of the shock \citep{Du2014,Du2015,Bhunia2023,Morosan2023,Zhang2024b} has been presented in several papers. In addition, \cite{Zimovets2015} studied a band-split type II burst showing a three-lane structure. They suggested that two of the three lanes agree with the upstream-downstream model, whereas the third lane originates from a distinctly separate location. However, previous studies had some limitations regarding the capabilities of radio imaging, for example, a lack of multi-frequency imaging, low spatial resolution, or the unavailability of polarisation measurements. Modern radio interferometric arrays can now be used to study the origin of band-splitting in detail by simultaneously combining multi-frequency and high resolution radio images with spectropolarimetric observations. 

    In this paper, we present the imaging and spectropolarimetric study of a band-split type II solar radio burst using the Low Frequency Array \citep[LOFAR;][]{vanHaarlem2013}. We used LOFAR images of the type II harmonic lane to track the locations of the radio sources corresponding to the two components of the split band. In addition, we used LOFAR's full Stokes dynamic spectra to estimate the degree of circular polarisation (dcp) for both components in order to determine their possible origins.

    \begin{figure}[h!]
        \centering
        \includegraphics[width=\linewidth]{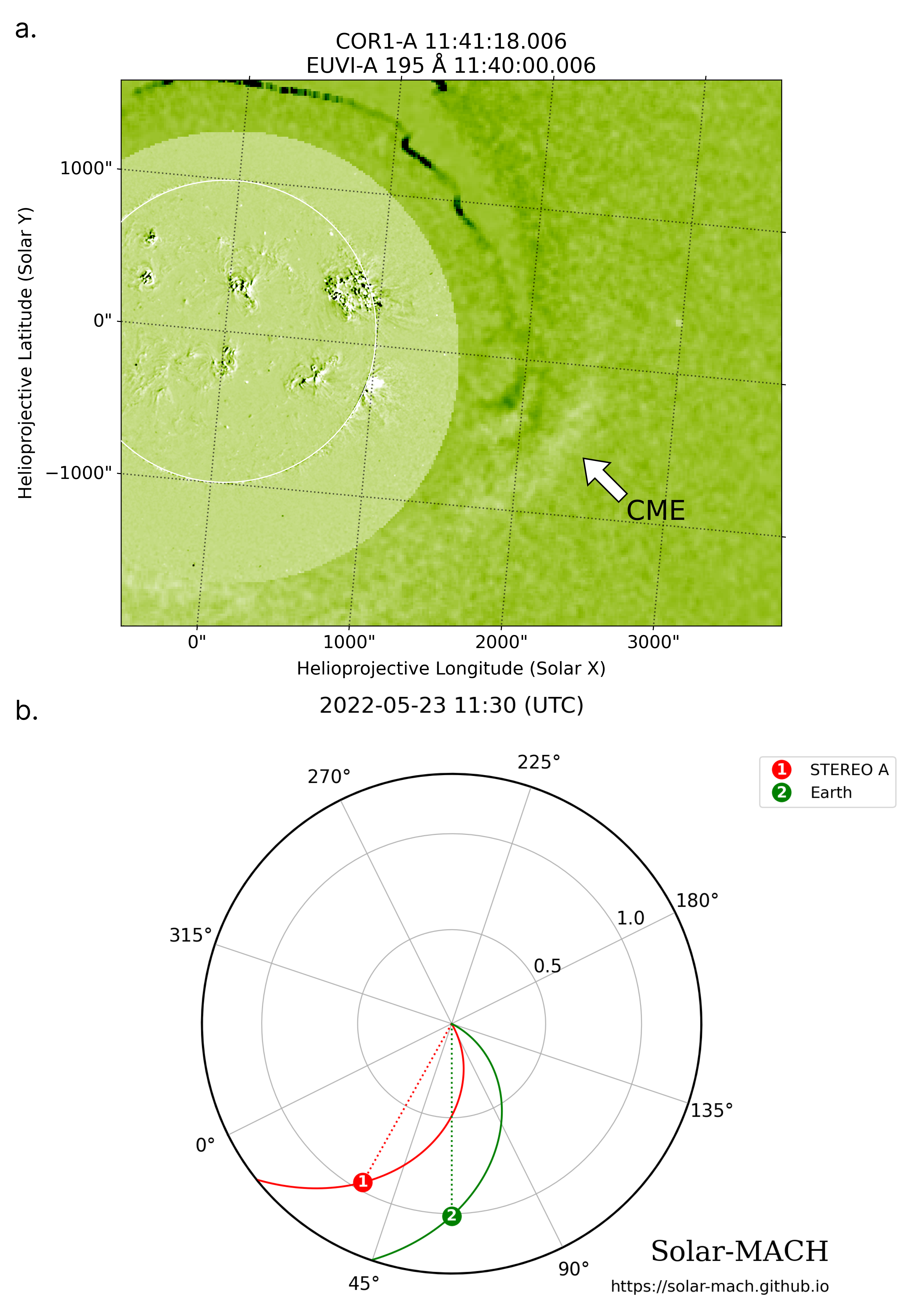}
        \caption{Coronal mass ejection associated with the type II burst on 23 May 2022 (see also Movie 1, available online) and the locations of the observing spacecraft. (a) Faint CME observed at $\sim$11:40 UT by STEREO-A. (b) Solar-MACH \citep{Gieseler2023} plot showing the location of STEREO-A (red) and Earth (green) at 11:30 UT. }
        \label{fig:1}
    \end{figure}
    
    \section{Observations and data analysis}

    \subsection{Radio observations}

\begin{figure*}[h!]
        \centering        
        \includegraphics[width=0.75\linewidth]{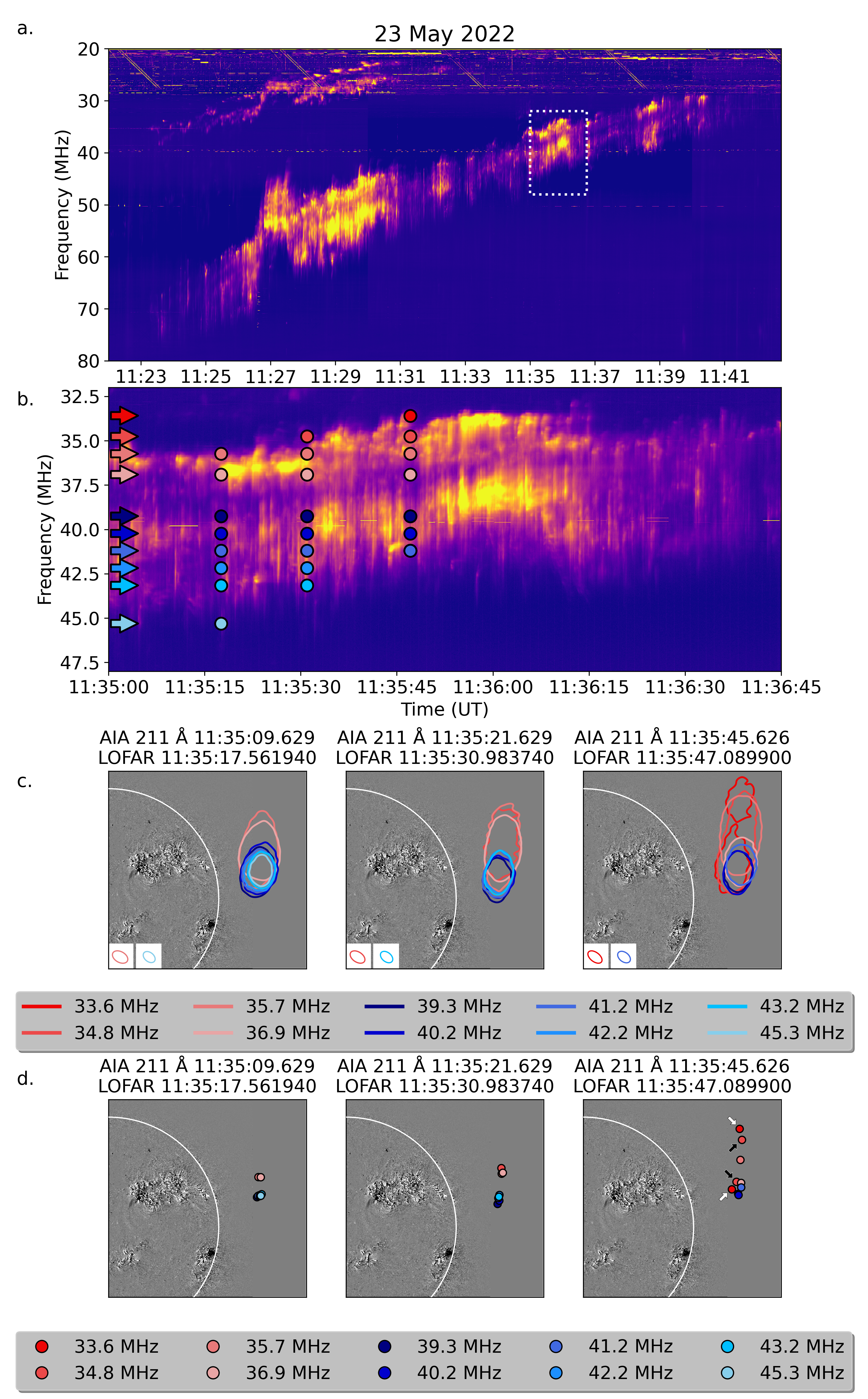}
        \caption{ Dynamic spectra and imaging of the type II solar radio burst observed by LOFAR on 23 May 2022. (a) Full dynamic spectrum of the type II burst between $\SI{20}{\mega\hertz}$ and $\SI{80}{\mega\hertz}$ at 11:22–11:42 UT. (b) Zoom-in of the spectrum outlined by a white dotted rectangle in (a). The coloured arrows indicate the central frequencies of the sub-bands used for the imaging. The coloured dots correspond to the three different time steps chosen for showing the imaging of the radio sources. (c) Contours of the radio sources overlaid on SDO/AIA $\SI{211}{\AA}$ running difference images. The contours were extracted from LOFAR imaging at $\sim$11:35:18 (left), $\sim$11:35:31 (middle), and $\sim$11:35:47 (right) UT. The times and colours correspond to the dots in (b). The temporal evolution of the radio contours between 11:35:00 and 11:36:45 UT is shown in Movie 3 (available online). (d) Centroids of the radio sources at the same frequencies and times as in (c). In the right panel, two sets of centroids were extracted for two frequencies and  are indicated by white ($\SI{33.6}{\mega\hertz}$) and black ($\SI{34.8}{\mega\hertz}$) arrows. }
        \label{fig:2}
    \end{figure*}

    On 23 May 2022, LOFAR observed a type II solar radio burst at approximately 11:23–11:41 UT. The type II burst was associated with a faint CME (see Fig. \ref{fig:1} and Movie 1 online) observed by the Solar Terrestrial Relations Observatory Ahead \citep[STEREO-A;][]{Kaiser2008}. The CME was observed at $\SI{195}{\AA}$ by the Extreme UltraViolet Imager \citep[EUVI;][]{Wuelser2004} and in white light by the Inner Coronagraph COR-1 \citep{Howard2008}. The dynamic spectrum of the type II burst is shown in Fig. \ref{fig:2}a at 11:22–11:42 UT between $\SI{20}{}-\SI{80}{\mega\hertz}$. The spectrum in Fig. \ref{fig:2}a and \ref{fig:2}b was observed by the low-band antennas (LBAs) from a single LOFAR core station (CS032) with a frequency resolution of $\SI{12.2}{\kilo\hertz}$ and a temporal resolution of $\SI{10}{\milli\second}$. The type II burst is composed of a fundamental and a harmonic lane, and each split into two sub-bands. In addition, after $\sim$11:27 UT, the type II burst shows a spectral bump that has recently been studied by \cite{Zhang2024b}. A white dotted rectangle in Fig. \ref{fig:2}a outlines the part of the spectrum seen in \ref{fig:2}b at 11:35:00-11:36:45 UT in the range of $\SI{32}{}-\SI{48}{\mega\hertz}$.
    
    In addition to the single station observations, data recorded by the LBAs in a total of 31 stations (23 core and 8 remote stations) were used in this study. The stations observed simultaneously in the interferometric imaging mode (31 stations) and in the beam-formed mode (core only). Both types of observations were recorded in 60 non-uniform frequency sub-bands in the range of $\SI{20}{}-\SI{80}{\mega\hertz}$. The total bandwidth of a single sub-band is $\SI{195.3}{\kilo\hertz}$ with a frequency resolution of $\SI{12.2}{\kilo\hertz}$. The imaging observations have a temporal resolution of $\SI{0.671}{\second}$. The beam-formed mode recorded Stokes $I$, $Q$, $U$, and $V$ dynamic spectra with a temporal resolution of $\SI{10.5}{\milli\second}$. Figure \ref{fig:3} shows the full Stokes dynamic spectra of the type II burst. These spectra were calibrated to solar flux units (sfu; $1 \, \mathrm{sfu} = \SI{e-22}{\watt\per\meter\squared\per\hertz}$) using observations of a calibrator (Cassiopeia A).

    \subsection{Processing of interferometric data}

    The processing of LOFAR's interferometric data presented here follows the methods from \cite{Zhang2024b}. The first step was to compute the gain solutions for each frequency sub-band and baseline using \verb|DP3| \citep[the Default Preprocessing Pipeline;][]{vanDiepen2018}. The gain solutions were computed by comparing the observation of the calibrator to its flux density model. Following this, the amplitude and phase plots for each antenna were inspected in case any unusable data needed to be flagged. For this dataset, none of the antenna stations were flagged. The gain solutions obtained in the first step were then applied to the observations of the Sun, resulting in a corrected phase and amplitude. In order to construct images, a two-dimensional inverse Fourier transform was applied to the observed visibilities. The final images were restored by applying the w-stacking CLEAN algorithm. Both of these actions were executed using \verb|wsclean| \citep{Offringa2014}. When running \verb|wsclean|, we used Brigg's weighting scheme \citep{Briggs1995} with a robustness parameter of $-0.5$. 

    \clearpage

    \clearpage

    \subsection{Correcting for polarisation leakage}

    Faraday rotation in the solar corona is expected cancel out any linear polarisation in solar radio bursts at low frequencies ($<\SI{100}{\mega\hertz}$) when observing over a certain bandwidth \citep[e.g.][]{Grognard1973,Morosan2022}. However, it is often the case that a linearly polarised signal, i.e. Stokes $Q$ and $U$, is observed in solar radio bursts at significant levels above the background. This is usually attributed to instrumental leakage from Stokes $I$ and $V$. We used the methods of \cite{Morosan2022} to correct the signals in Stokes $I$ and $V$ for leakage. The corrected Stokes $I$ and $V$ are defined as
    \begin{eqnarray}
        I_{corr} &=& I + F_Q \, I + F_U \, I \label{eq:Icorr} \\
        V_{corr} &=& \frac{V}{|V|} \sqrt{V^2 + (|U| - F_U \, I)^2} \label{eq:Vcorr},
    \end{eqnarray}
    where $F_Q$ and $F_U$ are the fraction of Stokes $I$ leakage to $Q$ and $U$, respectively. After inspecting the Stokes $I$, $Q$, $U$, and $V$ dynamic spectra, we found evidence of leakage from Stokes $I$ to both $Q$ and $U$. We also found evidence of leakage from Stokes $V$ to $U$. The leakage from Stokes $I$ into $Q$ and $U$ was calculated as $F_Q = |Q/I|$ and $F_U = |U/I|$, respectively. 

    The leakage from Stokes $I$ into $U$ was estimated using a part of the spectrum where no Stokes $V$ signal was present. We extracted the flux density time series from the spectra, and we averaged it over each frequency sub-band. A median background subtraction was also performed on the Stokes $I$ flux time series. We found a value of $F_U \approx 6\%$ and used this same value to correct Stokes $I$ at each time step and sub-band. Since we did not find evidence of Stokes $V$ leakage into $Q$, the leakage from Stokes $I$ into $Q$ was calculated for each time step and sub-band separately. The amount of leakage, $F_Q$, was on the order of a few percent, with values ranging from $0\%$ up to $6\%$. The evidence for the leakage from Stokes $V$ into $U$ was supported by the fact that the signal in Stokes $U$ was slightly stronger when Stokes $V$ was present compared to the instances with no Stokes $V$ signal.

    \section{Results}

    \subsection{Locations of the band-split radio sources}

    To determine the emission region of split bands, we investigated the location and polarisation of radio sources composing the type II harmonic band. Figure \ref{fig:2}b shows a zoomed-in dynamic spectrum of the harmonic band at 11:35:00–11:36:45 UT between $\SI{32}{\mega\hertz}$ and $\SI{48}{\mega\hertz}$. This part of the spectrum is outlined by a white dotted rectangle in the full dynamic spectrum in Fig. \ref{fig:2}a. We chose this part of the type II burst since it shows the highest signal in Stokes $V$ simultaneously with the least leakage in Stokes $Q$. Figure \ref{fig:2}b clearly shows the band-splitting as well as the herringbones composing the type II. In the spectrum, the coloured arrows indicate the centre frequencies of the ten sub-bands of which we have imaging available between $\SI{33.6}{\mega\hertz}$ and $\SI{45.3}{\mega\hertz}$. The colours for the different frequencies are chosen such that the red frequencies roughly correspond to the LFC and the blue colours to the HFC of the split band. The coloured dots in the spectrum correspond to the three time steps chosen for showing examples of the imaging. The radio images between 11:35:00 and 11:36:45 UT are available online as Movie 2. 

    We then extracted contours from the LOFAR imaging at the three time steps indicated in Fig. \ref{fig:2}b. Figure \ref{fig:2}c shows the radio contours at 11:35:18 (left), 11:35:31 (middle), and 11:35:47 UT (right). The contours outline the bright radio emission at $70\%$ of the maximum intensity in each image. The contours are plotted over running difference images of the Sun at $\SI{211}{\AA}$ observed by the Atmospheric Imaging Assembly \citep[AIA;][]{Lemen2012} on board the Solar Dynamic\textcolor{Green}{s} Observatory \citep[SDO;][]{Pesnell2012} at 11:35:10 (left), 11:35:22 (middle), and 11:35:46 UT (right). The radio contours in Fig. \ref{fig:2}c show the locations of the radio sources corresponding to the LFC (red) and the HFC (blue). The images show the presence of at least two distinct radio sources and are thus in agreement with the study of \cite{Zhang2024b}, which focused on the spectral bump in this type II burst. However, the study of \cite{Zhang2024b} was only performed at two frequencies (i.e. the frequency corresponding to each band). In the present study, we imaged all available sub-bands over the extent of the type II burst and found two separate frequency clusters (reds and blues). The blue frequencies (HFC) are clustered in one location, while the red frequencies (LFC) are clustered at another location. It should be also noted that the LFC radio sources appear more extended than the HFC radio source. It is possible that at the red frequencies there are actually two radio sources, but most of the time they cannot be easily resolved. This is further supported by the right panel in Fig. \ref{fig:2}c, where the contour at $\SI{33.6}{\mega\hertz}$ is split into two sources. Additionally, the contour at $\SI{34.8}{\mega\hertz}$ has a non-elliptical shape that also points to two sources being present. The separation of the LFC and the HFC radio sources is rather consistent throughout the time range from 11:35:00 to 11:36:45 UT, as can been seen from the movie related to Fig. \ref{fig:2} accompanying this paper (Movie 3). 

    The separation of the LFC and the HFC radio sources becomes even more evident when centroids of the radio contours in Fig. \ref{fig:2}c are extracted. This is done by fitting a two-dimensional elliptical Gaussian function to the LOFAR imaging and finding the peak of the function. Figure \ref{fig:2}d shows the centroids of the radio sources seen in Fig. \ref{fig:2}c. The centroids are overlaid on running difference images of the Sun at $\SI{211}{\AA}$ observed by SDO/AIA. The times of the EUV images and the centroids are the same as in Fig. \ref{fig:2}c. Again, the left and middle panels especially show two clearly separated groups of centroids: reds, corresponding to the LFC, and blues, corresponding to the HFC. In the right panel, taking into account the two sources at $\SI{33.6}{\mega\hertz}$ and at $\SI{34.8}{\mega\hertz}$, two sets of centroids were extracted at the corresponding frequencies as opposed to the typical treatment of one centroid per frequency. These two sets are marked with white ($\SI{33.6}{\mega\hertz}$) and black ($\SI{34.8}{\mega\hertz}$) arrows. Although in this instance there are some similar locations between the blue centroids and some of the red centroids, there are still a few centroids at the red frequencies that are clearly separated. In Fig. \ref{fig:2}d, the distance between the centroids at $\SI{35.7}{\mega\hertz}$ and $\SI{40.2}{\mega\hertz}$ in each panel is $\SI{114}{\mega\meter}$ (left), $\SI{168}{\mega\meter}$ (middle), and $\SI{219}{\mega\meter}$ (right).
    
    \begin{figure*}
        \centering
        \includegraphics[width=0.9\linewidth]{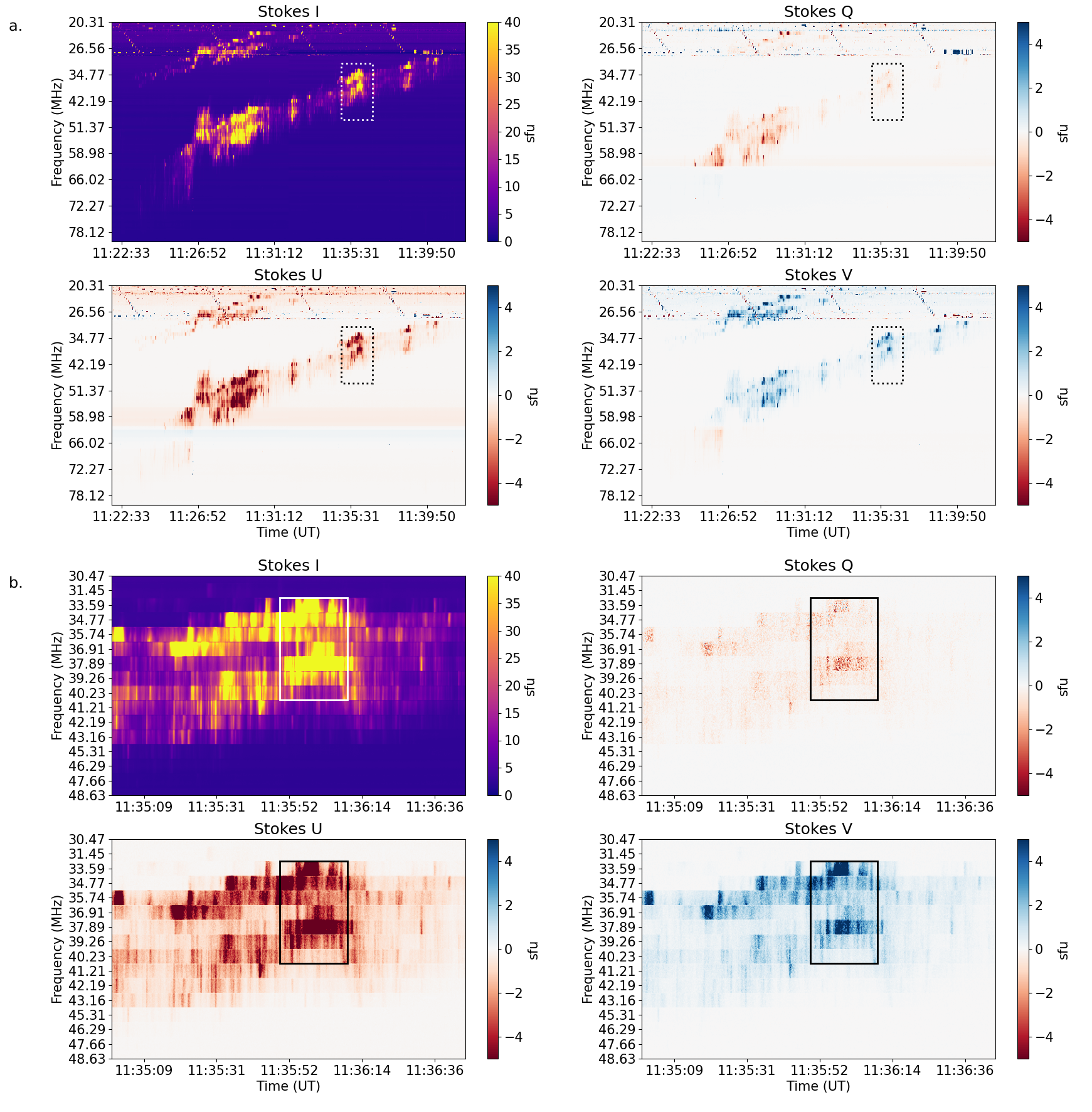}
        \caption{ Full Stokes dynamic spectra of the type II burst observed by LOFAR. In both (a) and (b), the Stokes $I$ (top left), $U$ (bottom left), $Q$ (top right), and $V$ (bottom right) are displayed. (a) Dynamic spectrum in the range of approximately $20-\SI{80}{MHz}$ at 11:22–11:42 UT. The dashed rectangle in each panel outlines the part of the spectra shown in (b). (b) Zoomed-in parts of the dynamic spectra in the range at 11:35:00–11:36:45 UT. The rectangles in each panel outline the part of the spectrum with the highest amount of the Stokes $V$ signal and the least leaked signal in Stokes $Q$. }
        \label{fig:3}
    \end{figure*}

    \begin{figure*}
        \centering
        \includegraphics[width=\linewidth]{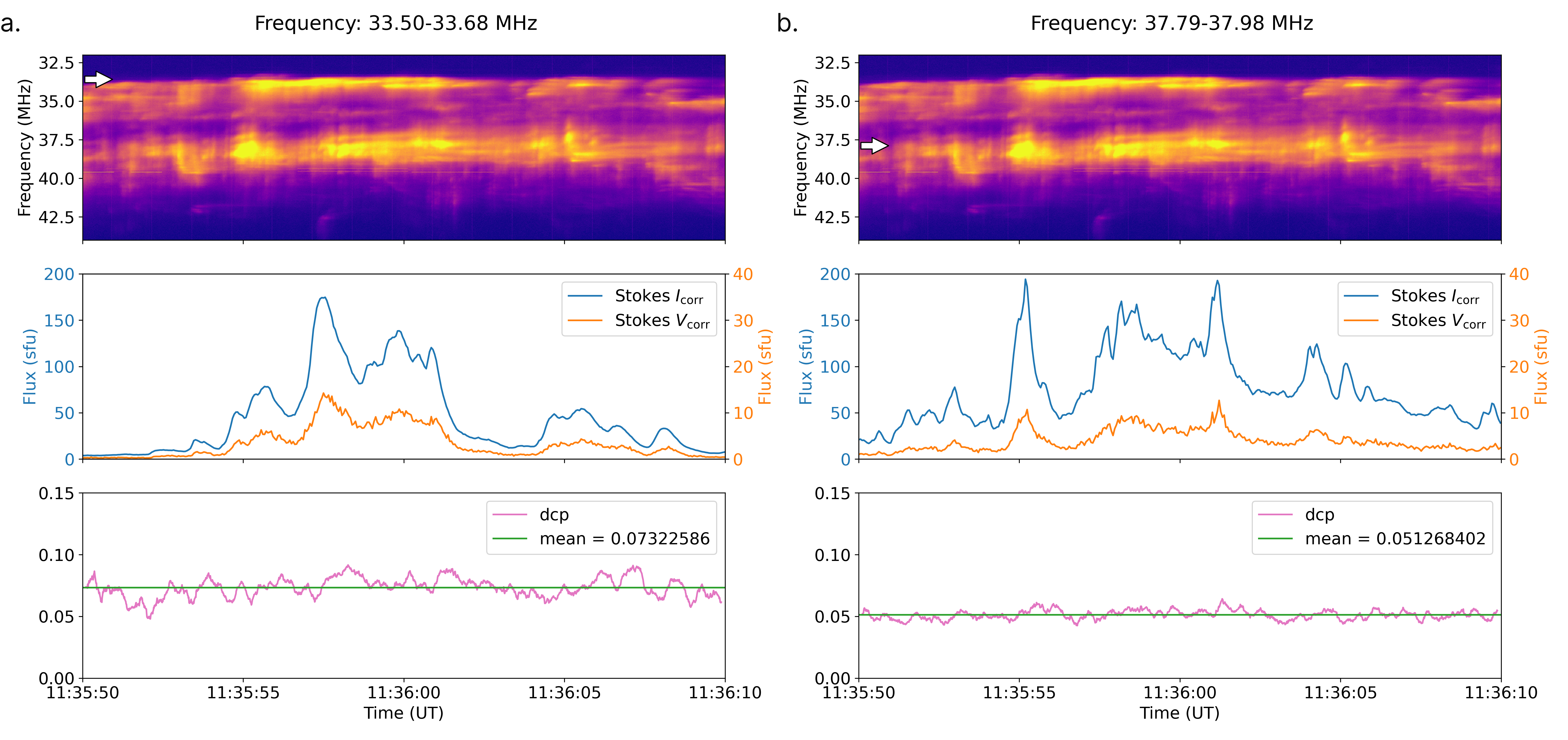}
        \caption{ Stokes $I$ and $V$ flux densities together with the dcp time series at $\SI{33.6}{\mega\hertz}$ (a) and at $\SI{37.9}{\mega\hertz}$ (b). The top panels in (a) and (b) show the dynamic spectrum at 11:35:50-11:36:10 UT in the range of $\SI{32}{}-\SI{44}{\mega\hertz}$. The white arrows indicate the frequencies of the used sub-bands. The middle panels show the time series of the corrected Stokes $I$ (blue) and $V$ (orange) flux densities. The flux densities are averaged over $\SI{0.05}{\second}$. The bottom panels show the time series of the dcp (pink) and its mean value (green). The dcp time series has been smoothed using the centred moving average method (window size: 21). }
        \label{fig:4}
    \end{figure*}

    \subsection{Degree of circular polarisation}

    The full Stokes dynamic spectra observed by LOFAR is shown in Fig. \ref{fig:3}. In both Fig. \ref{fig:3}a and \ref{fig:3}b, the Stokes $I$ (top left), $U$ (bottom left), $Q$ (top right), and $V$ (bottom right) are shown. Figure \ref{fig:3}a shows the full type II burst at 11:22–11:42 UT between approximately $\SI{20}{\mega\hertz}$ and $\SI{80}{\mega\hertz}$. Figure \ref{fig:3}b shows zoomed-in parts of the spectra at 11:35:00–11:36:45 UT, which are outlined with dotted rectangles in Fig. \ref{fig:3}a. We note that the frequency coverage is not continuous for the full Stokes spectra, so the marks on the y-axis indicate the centre frequencies of the observed sub-bands. The solid rectangles in Fig. \ref{fig:3}b outline the part of the spectrum where there is the highest amount of the Stokes $V$ signal in both components of the split band. Thus, we used this part of the spectrum to estimate the dcp for the LFC and the HFC.

    We calculated the dcp as
    \begin{eqnarray}
        dcp &=& \frac{V_{corr}}{I_{corr}} \label{eq:dcp},
    \end{eqnarray}
    where $I_{corr}$ and $V_{corr}$ are the corrected Stokes $I$ (Eq. \ref{eq:Icorr}) and Stokes $V$ (Eq. \ref{eq:Vcorr}) flux densities, respectively. We calculated the dcp time series by extracting and correcting the Stokes $I$ and $V$ fluxes from two frequency sub-bands within the rectangles in Fig. \ref{fig:3}b. The results are shown in Fig. \ref{fig:4}. The top panels in Fig. \ref{fig:4}a and \ref{fig:4}b show the zoomed-in dynamic spectrum in the range of $\SI{32}{}-\SI{44}{\mega\hertz}$ at 11:35:50–11:36:10 UT. The white arrows mark the centre frequencies of the used sub-bands: $\SI{33.6}{\mega\hertz}$ (Fig. \ref{fig:4}a) and $\SI{37.9}{\mega\hertz}$ (Fig. \ref{fig:4}b). The middle panels show the $I_{corr}$ (blue) and $V_{corr}$ (orange) flux time series. These particular sub-bands were chosen since they show the highest amount of the Stokes $V$ signal within this time period. The fluxes are averaged over $\SI{0.05}{\second}$ in the plot. However, we used the full time resolution data to calculate the dcp (Eq. \ref{eq:dcp}) time series. In addition, a median background subtraction was performed on the Stokes $I$ time series. The dcp time series (pink) and its mean value (green) are shown in the bottom panels of Fig. \ref{fig:4}a and \ref{fig:4}b. Here, the time series have been smoothed using the centred moving average method with a window size of 21 that corresponds to a time step of approximately $\SI{0.22}{\second}$. The results show that the mean dcp value is rather low for both cases: $\approx 7 \%$ for the LFC (Fig. \ref{fig:4}a) and $\approx 5 \%$ for the HFC (Fig. \ref{fig:4}b). The values are very similar for both cases, but our results indicate a slightly higher dcp for the LFC than for the HFC. All sub-bands corresponding to the LFC had mean dcp values in the range of approximately $6-7 \%$, whereas the sub-bands corresponding to the HFC all had values of approximately $5 \%$.
    
    The dcp can be used to estimate the strength of the local magnetic field as \citep[e.g.][]{Dulk1980,Ramesh2022a,Ramesh2022b}
    \begin{eqnarray}
        B \, [\mathrm{G}] &=& \frac{dcp \times f_p}{2.8 \, a(\theta)} \label{eq:B},
    \end{eqnarray}
    where $f_p$ is the local plasma frequency in units of $\SI{}{\mega\hertz}$, and 
    \begin{eqnarray}
        a(\theta) &=& \frac{16 + 11 \cos^2{\theta}}{48\cos{\theta}} \label{eq:a},
    \end{eqnarray}
    where $\theta$ is the angle between the magnetic field direction and the line of sight. Equation \ref{eq:B} shows that the dcp and the strength of the magnetic field $B$ are directly proportional to each other. The magnetic field ratio between the two bands can be calculated using the following relation:
    \begin{eqnarray}
        \frac{B_{LFC}}{B_{HFC}} = \frac{dcp_{LFC} \times f_{p,LFC}}{dcp_{HFC} \times f_{p,HFC}} \approx 0.8 \label{eq:Bd/Bu}.
    \end{eqnarray}
    If the two bands originate upstream and downstream of the shock, respectively, then $B_{LFC}$ would represent the upstream magnetic field, $B_{HFC}$ would be the downstream magnetic field \citep{Smerd1974,Smerd1975}, and Equation \ref{eq:Bd/Bu} would correspond to the shock's magnetic compression ratio. However, in this scenario, there is insignificant or no compression identified by this ratio. 
    
    \section{Discussion and conclusion}

    In this study, we used LOFAR observations to study the origin of band-splitting in a type II solar radio burst. We tracked the locations of the band-split radio sources over multiple frequencies and at multiple time steps using LOFAR imaging, and we found two clearly separated radio sources. In addition, we determined the dcp for the same time interval from LOFAR's full Stokes dynamic spectra, and we found similar values for both the HFC and the LFC. Both of these results support the \cite{Holman1983} model for band-splitting.

    In the \cite{Smerd1974,Smerd1975} band-splitting model, the two components of the emission band originate upstream and downstream of the shock. In this scenario, the two radio sources corresponding to the LFC and the HFC should have little to no separation in space. This model is contradicted by Fig. \ref{fig:2}c and \ref{fig:2}d, where we observed two distinct radio sources. The spatial separation is rather consistent throughout the imaging at 11:35:00–11:36:45 UT. A similar separation was found by \cite{Zhang2024b} when investigating the spectral bump of the type II burst between 11:25:00 and 11:30:30 UT. Other studies of band-split type II bursts have also found separation values in agreement with this study: \cite{Morosan2023} found a separation of $\SI{96}{\mega\meter}$, and \cite{Bhunia2023} found separations ranging from approximately $\SI{120}{}$ up to $\SI{372}{\mega\meter}$. The results obtained here therefore support the \cite{Holman1983} band-splitting model, where the components of the split band originate at the different locations upstream of the shock. On the other hand, \cite{Chrysaphi2018}, for example, found a spatial separation between the LFC and the HFC radio sources but explained the separation as being due to the scattering effects of radio waves \citep[see e.g.][]{Kontar2017,Kontar2019,Zhang2021}. However, we considered harmonic emission in this work, which is less affected by scattering than fundamental emission \citep[e.g.][]{Zhang2021}. In addition, \cite{Zhang2024b} found the separation between the LFC and the HFC radio sources at the beginning of this type II to be larger than the radial offset caused by propagation effects \citep{Zhang2021}. Further, in the present study, the clustering of sources at consecutive frequencies into two distinct regions cannot be explained by scattering. This is due to the fact that scattering is frequency dependent, and its effect would be to separate the multi-frequency sources equally instead of into clusters. Propagation effects are unlikely to be significant in this study.

    We found more supporting evidence for the \cite{Holman1983} model when estimating the dcp for the LFC and the HFC of the split band. The lower panels of Fig. \ref{fig:4}a and \ref{fig:4}b show that the mean dcp value for both components is low, with values of approximately $7\%$ (LFC) and $5\%$ (HFC). The values are very similar, but we found a slightly higher dcp for the LFC. However, this is not what we would expect if again considering the upstream-downstream scenario where the LFC originates upstream and the HFC originates downstream of the shock. We estimated a slightly higher dcp value for the LFC, which would imply a slightly higher magnetic field strength upstream of the shock. However, the upstream and downstream regions of the shock are two different plasma environments. The downstream plasma is more turbulent and compressed, which means higher values of magnetic field strengths with high amplitude fluctuations are expected \citep[e.g.][]{Bemporad2010,Kilpua2019}. This in turn would translate into higher values of dcp with significant fluctuations over time compared to the upstream region, which is not observed in our observation. In addition, the upper and lower frequency band magnetic ratio would correspond to a hypothetical magnetic compression ratio that is less than one, which is the opposite of what is expected of the magnetic ratio between the upstream and downstream shocked plasma. In addition, the obtained ratio does not match to the magnetic compression ratio for a quasi-perpendicular shock, which is expected to have a value similar to the density compression ratio \citep[e.g.][]{Kilpua2015}. For example, \cite{Bemporad2010} found a density and a magnetic compression ratio of $2$, and \cite{Simnett2005} found a density and a magnetic compression ratio of $3.7$. In addition, the results obtained in this study differ from the results of \cite{Ramesh2022b}, where a magnetic compression ratio of approximately two was estimated. However, \cite{Ramesh2022b} calculated the dcp for both components at the same frequency (i.e. at different times). This can be problematic since type II bursts propagate through different regions of the corona over time as the shock expands \citep[e.g.][]{Morosan2023,Morosan2024}, and thus the ambient plasma parameters are likely to change. The results in the present study reflect no significant changes in the magnetic environment during the emission of the two split bands composing the type II radio burst.

   There is further evidence that disagrees with the upstream-downstream band-splitting model. As already mentioned, the production of type II bursts requires the excitation of Langmuir waves \citep{Mann1995}. Langmuir waves have been observed only upstream of interplanetary shocks \citep[e.g.][]{Gurnett1979,Thejappa2000,Pulupa2010, Graham2015} and at the Earth's bow shock \citep[e.g.][]{Lacombe1985}. In addition, there is a disagreement between the Alfvén Mach numbers derived from the required Langmuir wave generation and from the upstream-downstream interpretation of type II band-splitting. Shock drift acceleration at a nearly perpendicular shock is able to accelerate electrons that excite Langmuir waves, which are necessary for the generation of type II bursts \citep{Holman1983,Mann2018}. \cite{Mann2022} found that electrons accelerated by shock drift acceleration excite Langmuir waves efficiently when the Alfvén Mach number of the shock is $1.59<M_A<2.53$ for coronal plasma conditions at the $\SI{25}{\mega\hertz}$ level. The maximum growth rate is obtained when $M_A=1.94$. However, previous studies using the upstream-downstream interpretation to determine the Alfvén Mach number have found rather low values: $M_A=1.06-1.16$ \citep{Zimovets2012}, $M_A=1.231-1.416$ \citep{Mahrous2018}, $M_A=1.3-1.5$ \citep{Zucca2018}, and $M_A=1.45-1.78$ \citep{Kumari2019}. The results of \cite{Mann2022} suggest that very low Alfvén Mach number shocks are not capable of accelerating electrons to energies sufficient for exciting Langmuir waves, which are essential for type II burst generation.
     
\begin{acknowledgements}{S.N. and D.E.M. acknowledge the Research Council of Finland project `SolShocks' (grant number 354409). This study has received funding from the European Union’s Horizon Europe research and innovation programme under grant agreement No.\ 101134999 (SOLER). The paper reflects only the authors' view and the European Commission is not responsible for any use that may be made of the information it contains. The research is performed under the umbrella of the Finnish Centre of Excellence in Research of Sustainable Space (FORESAIL) funded by the Research Council of Finland (grant no. 352847). The authors wish to acknowledge CSC – IT Center for Science, Finland, for computational resources. This paper is based on data obtained with the LOFAR telescope (LOFAR-ERIC) under project code \verb|LT16_001|. LOFAR (van Haarlem et al. 2013) is the Low Frequency Array designed and constructed by ASTRON. It has observing, data processing, and data storage facilities in several countries, that are owned by various parties (each with their own funding sources), and that are collectively operated by the LOFAR European Research Infrastructure Consortium (LOFAR-ERIC) under a joint scientific policy. The LOFAR-ERIC resources have benefited from the following recent major funding sources: CNRS-INSU, Observatoire de Paris and Université d'Orléans, France; BMBF, MIWF-NRW, MPG, Germany; Science Foundation Ireland (SFI), Department of Business, Enterprise and Innovation (DBEI), Ireland; NWO, The Netherlands; The Science and Technology Facilities Council, UK; Ministry of Science and Higher Education, Poland. } 
\end{acknowledgements}
    
    \bibliographystyle{bibtex/aa}
    \bibliography{references}

\end{document}